\documentclass{elsart}

\usepackage{graphicx}
\usepackage{amssymb}
\usepackage{amsmath}
\usepackage{rotating}
\begin{document}

\begin{frontmatter}



\title{Transport Induced by Mean-Eddy
 Interaction: \\
 II. Analysis of Transport Processes}
%  \thanksref{label1}}
%\thanks[label1]{Version 1}
\author[UMD]{Kayo Ide}\ead{ide@umd.edu}
\ead[url]{http://www.atmos.umd.edu/~ide}
 and
\author[UOB]{Stephen Wiggins} \ead{S.Wiggins@bris.ac.uk}
\ead[url]{http://www.maths.bris.ac.uk/people/faculty/maxsw/}
%\thanks[label2]{}
%\corauth[cor1]{}
\address[UMD]{Department of Atmospheric and Oceanic Science, \\
Center for Scientific Computation and Mathematical Modeling, \\
Institute for Physical Science and Technology, \\ \&
Earth System Science Interdisciplinary Center, \\
University of Maryland, College Park, USA}
\address[UOB]{School of Mathematics, University of Bristol, Bristol BS8 1TW, UK}
%\thanks[label3]{}

% abstract
\begin{abstract}
% Text of abstract
 We present a framework for the analysis of transport processes  resulting from
 the mean-eddy interaction in a flow.
 The framework is based on the {\bf T}ransport {\bf I}nduced by the  
 {\bf M}ean-{\bf E}ddy {\bf I}nteraction
 (TIME) method 
 presented in a companion paper \cite{ide_wiggins_pd06a}.
 The TIME method estimates the (Lagrangian) transport
 across stationary (Eulerian) boundaries  defined by chosen streamlines of
 the mean flow.
 Our framework proceeds after first carrying out a sequence of 
 preparatory steps that link the flow
 dynamics to the transport processes. 
 This includes the construction of the so-called
 ``instantaneous flux'' as the Hovm\"{o}ller diagram.
 Transport processes are studied by linking the signals of the
 instantaneous flux field   to the dynamical variability of the flow.
 This linkage also reveals how the variability of the flow contributes to the
 transport. The spatio-temporal analysis of the flux diagram can be used to assess 
 the efficiency of the variability in transport processes.
 We apply the method to the double-gyre ocean circulation 
 model in the situation  where the Rossby-wave mode dominates the dynamic variability. The spatio-temporal
 analysis shows that the inter-gyre transport is
 controlled by the circulating eddy vortices in the fast
 eastward jet region,
 whereas the basin-scale Rossby waves have very little impact.
\end{abstract}

\end{frontmatter}

% toc
\newpage
\renewcommand{\baselinestretch}{0.9}
\tableofcontents
\renewcommand{\baselinestretch}{1}
\newpage

% macro
\newcommand{\fginsrt}[1]{\marginpar{[Fig.\ref{#1}]}}
\newcommand{\tblinsrt}[1]{\marginpar{[Tab.\ref{#1}]}}

% main text
% section 1: Intro
\section{Introduction}
\label{sec:intro}

Analysis of  geophysical flows often employs
techniques that decompose the  velocity field in a manner that will
yield a desired insight \cite{peixoto_oort_book92}. A commonly
used technique is the mean-eddy decomposition

\begin{subequations}\label{eq:dyn}
\begin{eqnarray}
{\bf u}({\bf x},t)&=&
\overline{{\bf u}}({\bf x})+{\bf u}'({\bf x},t)~,\label{eq:dyn_u} \\
Q({\bf x},t)&=&\overline{ Q}({\bf x})+Q'({\bf x},t)~,\label{eq:dyn_q} 
\end{eqnarray}
\end{subequations}

\noindent
where the field is described by the unsteady eddy activity around a
mean state; henceforth, $\overline{ \{\cdot\}}$ and ${\{\cdot\}}'$ denote the
time average (mean) and the residual (unsteadiness or eddy),
respectively. Here ${\bf u}=(u_{1},u_{2})^{T}$ is the velocity field 
and  $Q$ represents any property of the flow such as temperature,
chemical or biological properties. In the absence of unsteadiness, the
kinematic transport occurs only along the mean streamlines on
which $\overline{{\bf u}}({\bf x})$ is  everywhere tangent.
An effect of the unsteadiness is to stir the flow instantaneously and induce mixing
across the mean streamlines over time.
It is also well accepted that instantaneous pictures of
the unsteady  flow themselves do not indicate transport explicitly.

A variety of Eulerian and Lagrangian methods have been developed
to study transport observationally, analytically, and numerically.
The companion paper \cite{ide_wiggins_pd06a} presented a new
transport method that is a hybrid combination of  Lagrangian and Eulerian
methods. This paper develops a framework for the analysis and
diagnosis of  transport processes based on this new method.
Our main focus  is on two-dimensional geophysical flows that may be
compressible. To introduce our method, we begin with a brief discussion of Lagrangian
and Eulerian methods specific to our needs.

The basis of any Lagrangian method involves the tracking of
individual fluid particles by solving the initial value problem of
$\frac{d}{dt}{\bf x}={\bf u}({\bf x},t)$. Starting from ${\bf x}_{0}$ at
time $t_{0}$, a particle trajectory ${\bf x}(t;{\bf x}_{0},t_{0})$  at
time $t$ is given by the temporal integral of the local velocity field
along itself: 

\begin{eqnarray} \label{eq:xL}
{\bf x}(t;{\bf x}_{0},t_{0})-{\bf x}_{0}&=&\int_{t_{0}}^{t}
{\bf u}({\bf x}(\tau;{\bf x}_{0},t_{0}),\tau)d\tau~. 
\end{eqnarray}

\noindent
A very rudimentary description of Lagrangian transport may be obtained
from a so-called ``spaghetti diagram'' which is constructed by simply plotting the
trajectories.  Typically this results in a complex tangle of curves from which detailed a detailed assessment of Lagrangian transport may prove difficult.  In recent  years the mathematical theory of dynamical systems has provided a new point of view and tools for classifying, organizing, and analyzing detailed and complex trajectory information by providing a theoretical and computational framework for an understanding of the geometric properties of
``flow structures''.  Recent reviews of the dynamical systems approach to transport are  given in \cite{wiggins_arfm05,msw,samwig}.

Nevertheless, there still remain many challenging problems to be
tackled by Lagrangian methods. Quantifying Lagrangian transport is
extremely elaborate in general. While techniques based on
dynamical systems theory are conceptually ideal for tracking
transport of fluid particles, they have not proven  as useful for studying
transport of $Q$, unless $Q$ is uniform and passive. 
Moreover,  Lagrangian
methods are not suitable for separating and/or isolating the roles
played by the mean state $\overline{{\bf u}}({\bf x})$ and 
the unsteadiness ${\bf u}'({\bf x},t)$ in the trajectory
${\bf x}(t;{\bf x}_{0},t_{0})$.

In contrast to  Lagrangian methods, transport quantities computed
with Eulerian methods utilize information taken at pre-selected
stationary points. At a station $\overline{{\bf x}}^{E}$, the most basic
Eulerian transport may be given by the temporal integral of the
local velocity during a time interval $[t_{0},t_{1}]$: 
\begin{eqnarray}
\label{eq:xE} \int_{t_{0}}^{t_{1}} {\bf u}(\overline{{\bf x}}^{E},t) dt &=&
(t_{0}-t_{1})\overline{{\bf u}}(\overline{{\bf x}}^{E})~. 
\end{eqnarray} 
The resulting transport is associated with 
$\overline{{\bf u}}({\bf x})$, but not ${\bf u}'({\bf x},t)$
by default
[compare with (\ref{eq:xL}) for the Lagrangian case]. 
For
transport across a stationary Eulerian curve $E=\{\overline{{\bf x}}^{E}(p)\}$
where $p$ is a parameter along $E$, the total transport during
$[t_{0},t_{1}]$ over a spatial segment $[p_{A},p_{B}]$ is 
\begin{eqnarray} \label{eq:mE} 
\int_{t_{0}}^{t_{1}}\int_{p_{A}}^{p_{B}}
\frac{d}{dp}\overline{{\bf x}}^{E}(p)
\wedge  {\bf u}(\overline{{\bf x}}^{E}(p),t)~dpdt 
&=&
(t_{1}-t_{0})\int_{p_{A}}^{p_{B}} \frac{d}{dp}\overline{{\bf x}}^{E}(p) 
\wedge \overline{{\bf u}}(\overline{{\bf x}}^{E}(p))~dp~. 
\end{eqnarray} 
% Once again,  this Eulerian method
% suppresses the effects of ${\bf u}'({\bf x},t)$.
An advantage of the Eulerian methods is the ability to compute the
transport of $Q$ as well, by replacing ${\bf u}({\bf x},t)$ with
$Q({\bf x},t){\bf u}({\bf x},t)$. 
Overtime, the Eulerian methods  gives 
$\overline{ Q({\bf x},t) {\bf u}({\bf x},t)}=\overline{ Q}({\bf x})
\overline{{\bf u}}({\bf x})+ 
\overline{ Q'({\bf x},t) {\bf u}'({\bf x},t)}$. 
Hence, the Eulerian methods account for the statistical
contribution to transport at the second-order.

Our new transport method \cite{ide_wiggins_pd06a} has the unique
ability to identify the effects of the mean-eddy interaction
in a way that neither Lagrangian nor Eulerian
methods have accomplished. This advantage that comes from 
blending of the Lagrangian and Eulerian approaches. The method
uses information on a stationary (Eulerian) boundary curve $C$ to
estimate (Lagrangian) transport of both fluid particles and $Q$
across $C$ without requiring tracking of individual fluid particles. 
The method estimates the transport
by integrating the instantaneous effects of the
unsteady flux while taking the particle advection of the        
mean flow into account. We refer to our method that quantifies  the {\bf T}ransport {\bf I}nduced by the
{\bf M}ean-{\bf E}ddy interaction as ``TIME.''

\begin{figure}[!ht]
\begin{center}
\includegraphics[width=15.cm]{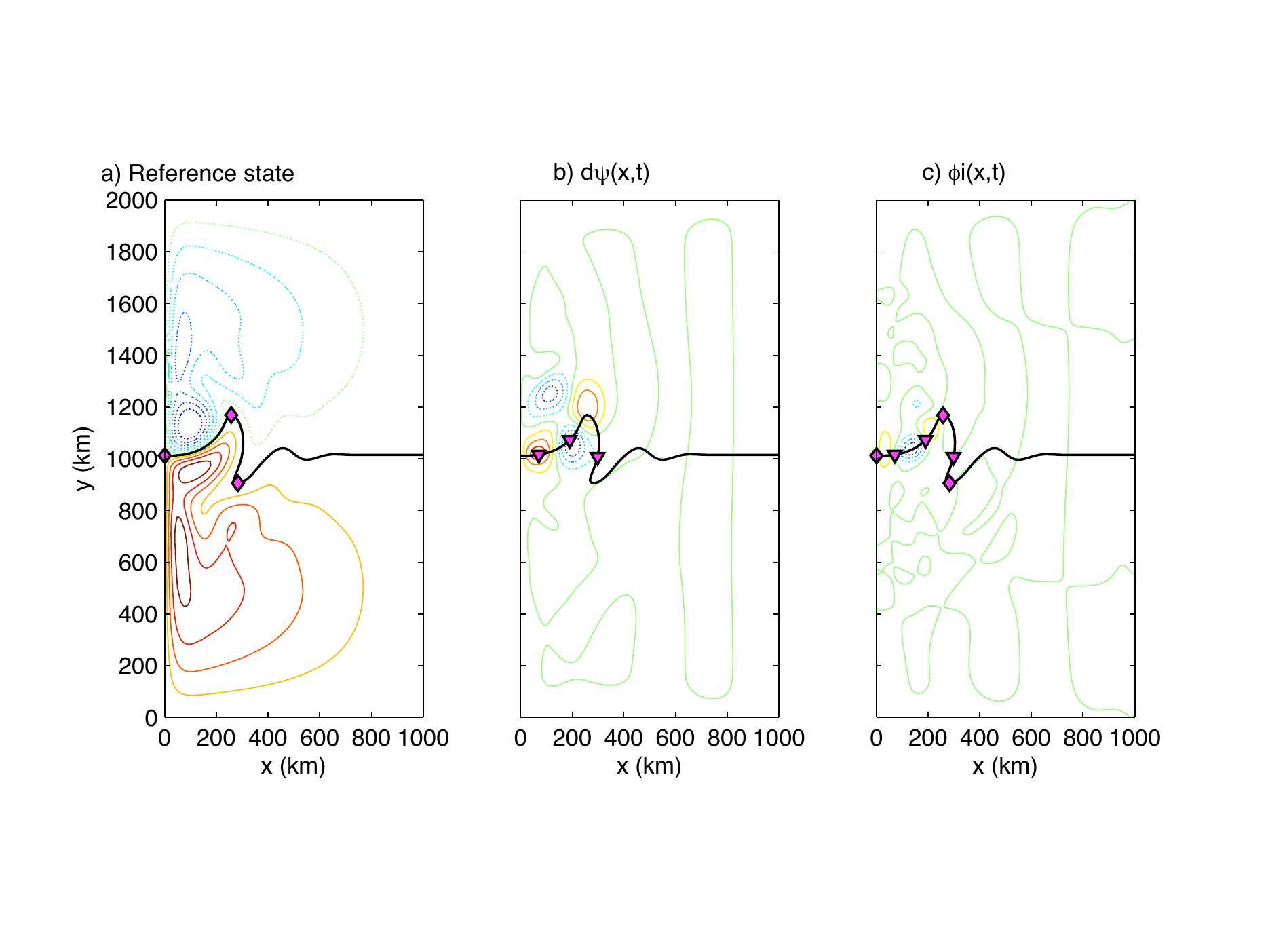}
\end{center}
\caption{Wind-driven double-gyre circulation at wind-stress curl
$0.165$dyn/cm$^{2}$: a) meaan streamline field $\overline{ \psi}({\bf x})$
averaged over $T$ with contour interval 2000; b) eddy streamline
field $\psi'({\bf x},t)$ at $t=t^{*}_{944}+T/4(=t^{*}_{982})$ 
with contour interval 500;
and c) instantaneous flux field
$\phi({\bf x},t)$ at $t=t^{*}_{944}+T/4(=t^{*}_{982})$ 
 with contour interval 50.
 (see also Figure~\ref{fg:dg_ph4_psie_phi}).
In all panels,
the dashed contours correspond to the negative values in all panels
the Eulerian boundary $C$ for the inter-gyre transport
shown by a thick solid line  
In $\overline{ \psi}({\bf x})$, 
  ${\bf x}_{{\rm J}}$,  
  ${\bf x}_{{\rm N}}$, and ${\bf x}_{{\rm S}}$  are shown by the
 diamonds;
In $\psi\prime({\bf x},t)$ and $\phi({\bf x},t)$,
${\bf x}_{1}$, ${\bf x}_{2}$, and ${\bf x}_{{\rm 3}}$ are shown
by the triangles.
  The flight-time coordinates of 
  ${\bf x}_{{\rm J}}$, ${\bf x}_{1}$, ${\bf x}_{2}$, 
  ${\bf x}_{{\rm N}}$, ${\bf x}_{{\rm S}}$ and ${\bf x}_{{\rm 3}}$ 
   are $s_{{\rm J}=110}$, $s_{1}=114.5$, $s_{2}=118.5$, 
     $s_{{\rm N}}=129$, $s_{{\rm S}}=174.5$ and $s_{{\rm 3}}=150$ days.}
\label{fg:dg_me}
\end{figure}

By construction, the TIME method offers a framework for a detailed
analysis of the spatio-temporal structure of transport processes. The
goal of this paper is to present this framework through a study of
the inter-gyre transport processes in  a wind driven, three-layer quasi-geostrophic ocean model
(Figure~\ref{fg:dg_me}). 
Due to its relevance to the mid-latitude ocean
circulation, the dynamics of wind-driven double-gyre ocean models
have been actively studied from various points of view over the
last few decades (for review, see \cite{dijkstra_book05} and references therein).
\fginsrt{fg:dg_me}

Although the details of transport are highly dependent on the
dynamics of the flows, there are five common preparatory
steps for  the analysis using the TIME method. 
The initial two steps of the five concern obtaining an  understanding 
of the flow dynamics based on the mean-eddy decomposition 
(\ref{eq:dyn}). 
The first step is to examine the global flow structure given by the mean
flow $\overline{{\bf u}}({\bf x})$, which we call the 
\emph{reference state}
 (Figure~\ref{fg:dg_me}a for the
wind-driven ocean; see Section~\ref{sec:bsc_mean} for the details). 
The second step is to understand the nature of the unsteady eddy
 activity in ${\bf u}'({\bf x},t)$. 
Unsteady eddy activity is also referred to  as the 
\emph{variability}. In  geophysical flows, variability is often
associated with the temporal evolution of the spatially coherent
structures (Figure~\ref{fg:dg_me}b; see Section~\ref{sec:bsc_dyn} for
the details). It should be clear that these coherent structures are
defined in the instantaneous Eulerian field and different from the
so-called ``Lagrangian coherent structures''
 \cite{haller_pfa02}.

With understanding and insights of  the mean and the variability at hand, the third
step is to compute the the instantaneous flux that stirs the flow.
The instantaneous flux is  expressed naturally  in terms of a  ''mean-eddy interaction''.  
(Figure~\ref{fg:dg_me}c for the wind-driven ocean; see
Section~\ref{sec:bsc_flx}).

The fourth step is to select the \emph{Eulerian boundary}, $C$, of interest based on
the mean flow structure. 
Any mean streamline with reasonable length can be a
potential $C$. The actual choice of $C$ should be left up to the
specific geophysical interests . In Figure~\ref{fg:dg_me}a  we show our choice of $C$ for the intergyre
transport in the wind-driven ocean;  which we discuss further in Section~\ref{sec:bsc_C}).

The last step is to extract the information of
the instantaneous flux on $C$. The signals of the variability in
the  mean-eddy interaction are conveniently represented by
 a space-time diagram (i.e., \emph{the Hovm\"{o}ller diagram}), which we call
 the \emph{flux  diagram} to emphasize its role in the TIME method
 (see Section~\ref{sec:bsc_mu}).

An important outcome of the five preparatory steps is that they 
link the flow dynamics to the transport processes, 
and {\it vice versa}, in terms of the mean-eddy interaction. 
This link, and the spatio-temporal integration employed 
by the TIME method, comprise the foundation for the graphical approach to
the study of the transport processes (see Section~\ref{sec:fnc}).
In the double-gyre application, the analysis will reveal 
how/when/where the circulating eddy vortices in a
localized area over the eastward jet are responsible for the
inter-gyre transport, whereas the basin-scale Rossby waves play
a very small role (see Section~\ref{sec:dg}).
We use the same flow field as in  \cite{ide_wiggins_pd06a}
that is nearly periodic in time and has a heteroclinic
connection in the mean.
It is worth noting that the TIME method does not require either of such
conditions (i.e., presence of the time periodicity and the heteroclinic
connection).  

The outline of the paper is as follows. Section~\ref{sec:bsc} presents the
preparatory steps using the application to the inter-gyre transport in
the double-gyre wind-driven ocean circulation. We extend the TIME
functions defined by \cite{ide_wiggins_pd06a} and present a
graphical approach that facilitates the analysis of transport
processes in Section~\ref{sec:fnc}. Inter-gyre transport processes in the
double-gyre ocean are analyzed in
Section~\ref{sec:dg}. Section~\ref{sec:cncl} 
summarizes the results and provides a discussion.

% section 2: Model
\section{Building  the links between variability and transport}
\label{sec:bsc}

In this section we  define the five preparatory steps 
mentioned in the previous section in detail and carry them out
in the context of  the analysis of  inter-gyre transport 
in a wind driven double-gyre
ocean circulation model. The flow field is obtained by 
numerical simulation of a three-layer quasi-geostrophic model with
the model parameters chosen to be consistent with the mid-latitude,
wind-driven  ocean
circulations \cite{dijkstra_book05,simonnet_etal_jpo03}. As a
result of a constant wind-stress curl $0.165$dyn/cm$^{2}$ applied
at the ocean surface,
the basin-scale circulation fluctuates
almost periodically around the mean state with dominant spectral peak at
period $T\approx 151$days
after the initial 30,000-day spin-up from the rest.  For our analysis time 
starts after this spin-up.
In this study, we analyze the inter-gyre transport processes in the top
layer.
The instantaneous flow patterns are given by the streamfunction
$\psi({\bf x},t)$ that is related to the velocity 
${\bf u}({\bf x},t)$ by
${\bf u}({\bf x},t)
=(-\frac{\partial }{\partial y},
  ~\frac{\partial }{\partial x})\psi({\bf x},t)$.

\subsection{Mean flow}
\label{sec:bsc_mean}

Given $\overline{{\bf u}}({\bf x})$, the corresponding 
streamfunction $\overline{\psi}({\bf x})$  is the reference state that
provides a  geometrical structure of the global flow.
 Figure~\ref{fg:dg_me}a shows
$\overline{\psi} ({\bf x})$ for  the 
double-gyre circulation in which the axis of the eastward jet 
divides the ocean basin into two gyres.
Because the ocean model  has gone through the
pitchfork bifurcation prior to the Hopf bifurcation
\cite{dijkstra_book05} (in terms of increasing wind-stress curl),
an asymmetry exists between the cyclonic subpolar gyre and the
anticyclonic subtropical gyre. At the confluence of the southward
and northward western boundary currents, the cyclonic subpolar
gyre and anti-cyclonic subtropical gyre 
form an asymmetric dipole 
(along ${\bf x}_{{\rm J}}$ and ${\bf x}_{{\rm N}}$ 
indicated by the diamonds in Figure~\ref{fg:dg_me}a).
We refer to this region as the ``dipole region.''
The eastward jet defined between the center of the subpolar vortex
and that of the subtropical gyre
and carries the mean net transport 
$\overline{\psi}_{NT}
\equiv|\overline{\psi}_{{\rm sp}}-\overline{\psi}_{{\rm st}}|=24028$,
where  $\overline{\psi}_{{\rm sp}}$ ($<  0$) is the minimum 
$\overline{\psi}({\bf x})$
over the subpolar vortex and 
$\overline{\psi}_{{\rm st}}$ ($>  0$) is the maximum over the
subtropical vortex.
The asymmetry of the two gyres is measured by the transport difference,
\cite{chang_etal_jpo01}
$\overline{\psi}_{TD} 
\equiv|\overline{\psi}_{{\rm sp}}|-|\overline{\psi}_{{\rm tr}}|=5438$.

\subsection{Dynamic variability of the flow in ${\bf u}'({\bf x},t)$}
\label{sec:bsc_dyn}

An understanding of the eddy activity (variability) in
${\bf u}'({\bf x},t)$ is the key for the analysis of transport
processes using the TIME method.
In the double-gyre application,
the model ocean dynamics is almost periodic in time
resulting from  the Rossby-wave mode
\cite{dijkstra_book05,simonnet_etal_jpo03}.
In the left panel of Figure~\ref{fg:dg_must_u},
the evolution of the ocean dynamics is shown by two time series.
The net transport of the jet $NT(t)\equiv
|\psi_{{\rm sp}}(t)-\psi_{{\rm st}}(t)|/\overline{\psi}_{NT}$
normalized by the mean state net transport $\overline{\psi}_{NT}$ 
indicates the fluctuation of the jet strength around the mean.
The transport difference 
$TD(t)=
(|\psi_{{\rm sp}}(t)|-|\psi_{{\rm st}}(t)|)/\overline{\psi}_{TD}$
normalized by $\overline{\psi}_{TD}$
measures the fluctuation of the asymmetry
between the subpolar and subtropical gyres around the mean.
Here $\psi_{{\rm sp}}(t)$ is the minimum $\psi({\bf x},t)$ over the
subpolar gyre and $\psi_{{\rm st}}(t)$  is the maximum over the
subtropical gyre.
The two time series show that the amplitude of the fluctuation 
is of the order smaller than 0.1 with respect to the mean. This fact is important since the validity of the TIME functions requires the fluctuations to be small compared to the mean.

For convenience, we define the period of the $k$-th ocean oscillation 
by $T^{[k]}=[t^{*}_{38}+(k-1)T,t^{*}_{38}+kT)$
starting from $t^{*}_{38}$ when a minimum of $NT(t)$ occurs first time 
after  the spin-up;
the number in the subscript of $t^{*}$ represents time in
day from here on.
We also define $T^{[k.1]}$ and $T^{[k.2]}$ 
as the first and the second half of $T^{[k]}$, respectively.
During $T^{[k.1]}$, $NT(t)$ increases from  a minimum to a maximum
while $TD(t)$ reaches a minimum 
after about $T/4$.
Conversely during  $T^{[k.2]}$, $NT(t)$ decreases 
while $TD(t)$ reaches a maximum after about $3T/4$
 from the beginning of $T^{[k]}$.
\fginsrt{fg:dg_must_u}

The eddy streamfunction field $\psi'({\bf x},t)$ associated with
${\bf u}'({\bf x},t)$ provides an instantaneous 
pattern of the eddy activity. In the double-gyre application,
 $\psi'({\bf x},t)$ shows two types of eddy activity 
(Figure~\ref{fg:dg_me}b). One is the westward propagation of  Rossby
waves, whose latitudinally elongated structures are especially
visible in the eastern basin. Typically there are three waves in
the entire basin, although the one in the western basin is distorted
around the dipole region. The travel time of the wave from the
eastern to the western boundary is 453(=151$\times$3)days.
Each wave has the longitudinal width 333(=1000/3)km and travels
with the propagation phase speed  2.2(=1000/453)km/day.

As mentioned briefly in the introduction, for our study of inter-gyre transport in the
 wind-driven ocean, the axis of the eastward jet  is chosen to be $C$.
It is the separatrix that connects the western boundary to 
the eastern boundary.  We will discuss this in more detail in Section \ref{sec:bsc_C}.

\begin{figure}[!ht]
\begin{center}
\includegraphics[height=20.cm]{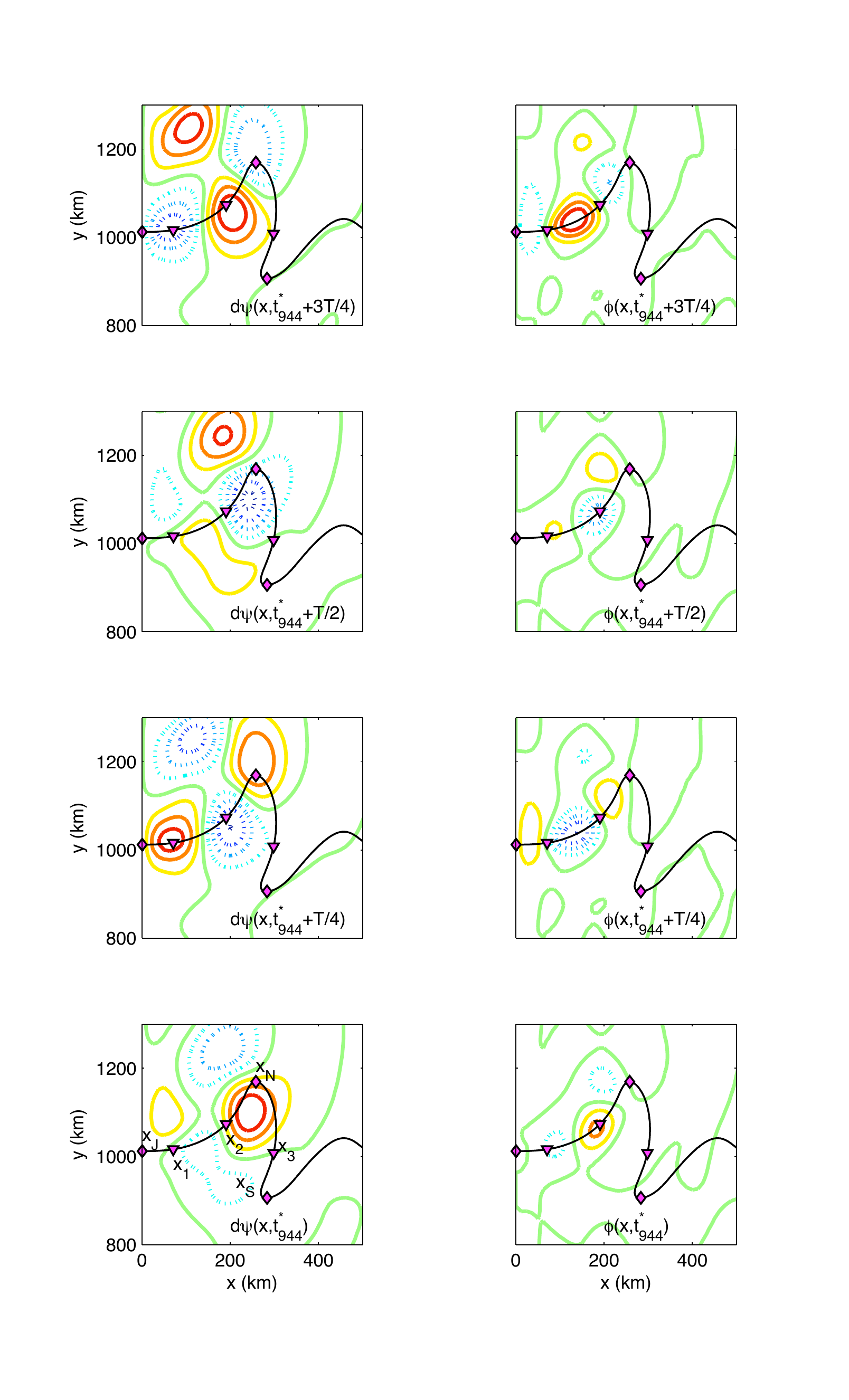}
\end{center}
\caption{The four phases of  $\psi'({\bf x},t)$ (left) and
$\phi({\bf x},t)$ (right) during $T^{[7]}$ 
at  $t^{*}_{944}$ when $NT(t)$ is minumum
$t^{*}_{944}+T/4(=t^{*}_{982})$ when $TD(t)$ is minimum,
$t^{*}_{944}+T/2(=t^{*}_{1020})$ when $NT(t)$ is maximum, and
$t^{*}_{944}+3T/4(=t^{*}_{1058})$ when $TD(t)$ is maximum
with time increasing upward.
The contour intervals are the same as in Figure~\ref{fg:dg_me}b for 
$\psi'({\bf x},t)$ and Figure~\ref{fg:dg_me}c for $\phi({\bf x},t)$.
The solid line with the diamonds and triangles is $C$:
from the upstream, the diamons show ${\bf x}_{{\rm J}}$, ${\bf x}_{{\rm N}}$, 
and  ${\bf x}_{{\rm S}}$;
the triangles show
${\bf x}_{1}$, ${\bf x}_{2}$, and  ${\bf x}_{{\rm 3}}$
as shown in Figure\ref{fg:dg_me}c.}
\label{fg:dg_ph4_psie_phi}
\end{figure}

Although the westward propagation of Rossby waves is seen in almost the
entire ocean basin, the dipole region has the eddy activity that is 
more energetic.
The left panels of Figure~\ref{fg:dg_ph4_psie_phi} show the phases of
these eddy vortices during $T^{[7]}$ that starts at 
$t^{*}_{944}$.
At $t^{*}_{944}+T/4(=t^{*}_{982})$ when $TD(t)$ is minimum 
(corresponding to Figure~\ref{fg:dg_me}b),
a strongly positive eddy vortex is located near ${\bf x}_{1}$.
This eddy vortex weakens as the center moves from ${\bf x}_{1}$ along
$C$ as shown at $t^{*}_{944}+T/2(=t^{*}_{1020})$ when
$NT(t)$ is maximum.
It intensifies once again as the center 
reaches  ${\bf x}_{2}$ 
as shown at $t^{*}_{944}+3T/4(=t^{*}_{1058})$  when $TD(t)$ is maximum.
It moves further along $C$ as shown at $t^{*}_{944}$ for $T$ later
($t^{*}_{944}+T=t^{*}_{1095})$,
until the center leaves $C$ near ${\bf x}_{{\rm N}}$
 as shown at $t^{*}_{982}$  for $5T/4$ later
($t^{*}_{944}+5T/4(=t^{*}_{1131})$).
And it continues to make a cyclic rotation around the sub-polar vortex.
One cyclic rotation  takes $2T$.
The negative eddy vortex located in the west of this positive
eddy vortex at $t^{*}_{944}+T/4$  follows the same cyclic motion 
but $T/2$ behind with the opposite phase of $NT(t)$ and $TD(t)$ 
with respect to the positive eddy vortex.
\fginsrt{fg:dg_ph4_psie_phi}

This cyclic rotation of the eddy vortices is far from uniform
because the eddy vortices tend to pulsate, i.e., 
when they intensify during $T^{[k.1]}$ or $T^{[k.2]}$
(e.g., $t^{*}_{944}+T/4$ or  $t^{*}_{944}+3T/4$ 
in Figure~\ref{fg:dg_ph4_psie_phi}), the centers
hardly move;
when they weaken at the end of $T^{[k.1]}$ or $T^{[k.2]}$
(e.g., $t^{*}_{944}$ or  $t^{*}_{944}+T/2$),
the centers move very quickly.
As we shall see below, this complexity in the flow dynamics 
strongly influences the inter-gyre transport processes.
We emphasize that these eddy vortices are not Lagrangian, i.e.,
particles don't move with them.

These  two types of eddy activity, westward propagation of the
Rossby wave and cyclonic circulation of the eddy vortices, are
synchronized. 
In particular they merge in the eastern part of the dipole region.
At $t^{*}_{944}+T/4$, a negative eddy vortex connects to
a negative half of the Rossby wave
around ${{\bf x}}_{3}$
as shown in Figures~\ref{fg:dg_me} and~\ref{fg:dg_ph4_psie_phi}.
This merger occurs every $T/2$  and is associated with the alternating sign of 
$\psi'({\bf x},t)$.

\subsection{Flux  variability of $\phi({\bf x},t)$}
\label{sec:bsc_flx}

The instantaneous flux across the mean flow

\begin{eqnarray} 
\phi({\bf x},t)&\equiv&
\overline{{\bf u}}({\bf x})\wedge{\bf u}'({\bf x},t)~\label{eq:phi_u} 
\end{eqnarray} 

\noindent
is explicitly defined in terms of the ``mean-eddy interaction'' induced
by the instantaneous spatial interaction of 
$\overline{{\bf u}}({\bf x})$  and ${\bf u}'({\bf x},t)$. 
Geometrically, $\phi({\bf x},t)$ is the signed area of
the parallelogram defined by $\overline{{\bf u}}({\bf x})$ and 
${\bf u}'({\bf x},t)$ with 
the unit of $\phi({\bf x},t)$ being flux per unit time over the length
$|\overline{{\bf u}}({\bf x})|$.
The amplitude of
$\phi({\bf x},t)$ depends not only $|\overline{{\bf u}}({\bf x})|$ 
and $|{\bf u}'({\bf x},t)|$, but also on the angle between 
$\overline{{\bf u}}({\bf x})$ and ${\bf u}'({\bf x},t)$.
We refer to the coherent structures associated with $\phi({\bf x},t)$ as the
\emph{flux zones}.

In the double-gyre application,
the two types of eddy activity in
${\bf u}'({\bf x},t)$ % or $\psi'({\bf x},t)$ 
lead to distinct evolution of flux zones in 
$\phi({\bf x},t)$. In the eastern basin where the Rossby waves dominate
the variability, the direction change of 
$\overline{{\bf u}}({\bf x})$ in both gyres 
breaks the wave structure latitudinally into three flux zones with
alternating signs (Figure~\ref{fg:dg_me}c). Near the jet
axis,  a sequence of the 
Rossby waves lead to a sequence of flux zones with alternating signs;
the positive flux zones correspond to northward flux, 
while the negative ones correspond to the southward flux.
The width and the westward  propagation  speed of
these flux zones are the same as those of the Rossby waves
in $\psi'({\bf x},t)$.

In the dipole region where $\overline{{\bf u}}({\bf x})$ is stronger 
and non-uniform while ${\bf u}'({\bf x},t)$ consists of the 
cyclic rotation of the four circulating eddy vortices, 
the mean-eddy interaction  
is complex (Figure~\ref{fg:dg_ph4_psie_phi}).
Most significantly, the  eddy vortices along the jet
lead to the flux zones with the alternating signs that are visible
particularly
where $|\overline{{\bf u}}(\overline{{\bf x}}^{C}(s))|$ is large
along the mean jet axis.
These flux zones  in $\phi({\bf x}, t)$ 
pulsate in sync with the eddy vortices in $\psi'({\bf x}, t)$
(compare the right panels of Figure~\ref{fg:dg_ph4_psie_phi} with the
left panels).
The shapes of the flux zones 
are more loosely defined than those of the eddy vortices due to
the spatially nonlinear  interference by $\overline{{\bf u}}({\bf x})$.
As the circulating eddy vortices intensify
when $TD(t)$ is minimum ($t^{*}_{944}+T/4$) or maximum ($t^{*}_{944}+3T/4$),
three relatively well-defined flux zones are visible over
$[{\bf x}_{{\rm J}}, {\bf x}_{1}]$, $[{\bf x}_{1},{\bf x}_{2}]$, 
and $[{\bf x}_{2},{\bf x}_{{\rm N}}]$  between the eddy vortex
centers with the middle one stronger than the other two.
In between consecutive intensifications, 
the flux zones weaken and  quickly propagate along $C$
as $NT(t)$ reaches minimum ($t^{*}_{944}$) or maximum
($t^{*}_{944}+T/2$).
The propagation speed and direction of the three flux zones
are the same as those of the  circulating eddy vortices.

As the eddy vortex leaves the mean jet axis near 
at ${\bf x}_{{\rm N}}$,
a weak  fourth  flux zone is generated over 
$[{\bf x}_{{\rm N}},{\bf x}_{3}]$ that propagate towards 
the upstream of the mean flow.
Because the eddy vortices and the Rossby waves merge around 
${\bf x}_{3}$ over $[{\bf x}_{{\rm N}},{\bf x}_{{\rm S}}]$, 
the flux zones associated with them also merge there.

\subsection{Boundary curve $C$ and parameterization of the reference
  trajectory}
\label{sec:bsc_C}

The TIME method uses the streamlines associated with
 $\overline{{\bf u}}({\bf x})$ as the Eulerian boundaries $C$ across
 which the transport is estimated. 
Along $C$, the flight-time $s$ is a natural choice of the coordinate
variable  because $C=\{\overline{{\bf x}}^{C}(s)\}$ is obtained by 
solving for $\frac{d}{ds}\overline{{\bf x}}^{C}(s)=
\overline{{\bf u}}(\overline{{\bf x}}^{C}(s))$ with a choice of initial
 position $\overline{{\bf x}}^{C}(0)$. 
Particle advection along $C$ in the mean flow is referred to as  the
\emph{reference trajectory}.
Starting from $\overline{{\bf x}}^{C}(s_{0})$ at $t_{0}$,
the reference trajectory is uniquely parameterized by $s_{0}-t_{0}$ and 
can be  written as
$(s,t)=(s_{0}-t_{0}+t,t)$ using the flight-time coordinate.

Some key locations on $C$ are shown in Figure~\ref{fg:dg_me}.
In $\overline{\psi}({\bf x})$,
${\bf x}_{{\rm J}}$ is the location that is far enough  
from the hyperbolic stagnation point of $C$ 
on the western boundary point  in $s$ so that  
$\overline{{\bf u}}({\bf x}_{{\rm J}})$ becomes
non-negligible to induce the instantaneous flux; 
${\bf x}_{{\rm N}}$ and ${\bf x}_{{\rm S}}$ 
are the locations  where the (meandering) jet axis makes 
sharp turns.
In $\psi'({\bf x},t)$, ${\bf x}_{1}$ and ${\bf x}_{2}$ 
are the locations where the centers of the eddy vortices 
pause to intensify;
${\bf x}_{{\rm N}}$ is where the eddy centers leave $C$;
around ${\bf x}_{3}$,
two types of variability,
the circulating eddy vortices in the upstream and 
the Rossby waves in the downstream, meet on $C$.
Accordingly,
${\bf x}_{{\rm J}}$, ${\bf x}_{1}$,
${\bf x}_{2}$, ${\bf x}_{{\rm N}}$ and ${\bf x}_{3}$ 
are the boundary points of the flux
zones along $C$ in  $\phi'({\bf x},t)$
as the eddy vortices pause to intensify and meet the
Rossby waves in ${\phi}({\bf x},t)$.
The flight-time coordinates of ${\bf x}_{{\rm J}}$,
${\bf x}_{1}$, ${\bf x}_{2}$, ${\bf x}_{{\rm N}}$, ${\bf x}_{3}$,
and ${\bf x}_{{\rm S}}$ are
$s_{{\rm J}}=110$, $s_{1}=114.5$, $s_{2}=118.5$, 
$s_{{\rm N}}=129$, $s_{3}=150$,
and $s_{{\rm S}}=174.5$ in the unit of days, respectively,
starting with
$\overline{{\bf x}}^{C}(0)=(2\times 10^{-38}{\rm km},1011.8{\rm km})$
located very close to the western boundary.

\subsection{Flux diagram}
\label{sec:bsc_mu}

\begin{figure}[!ht]
\begin{center}
\includegraphics[width=15.cm]{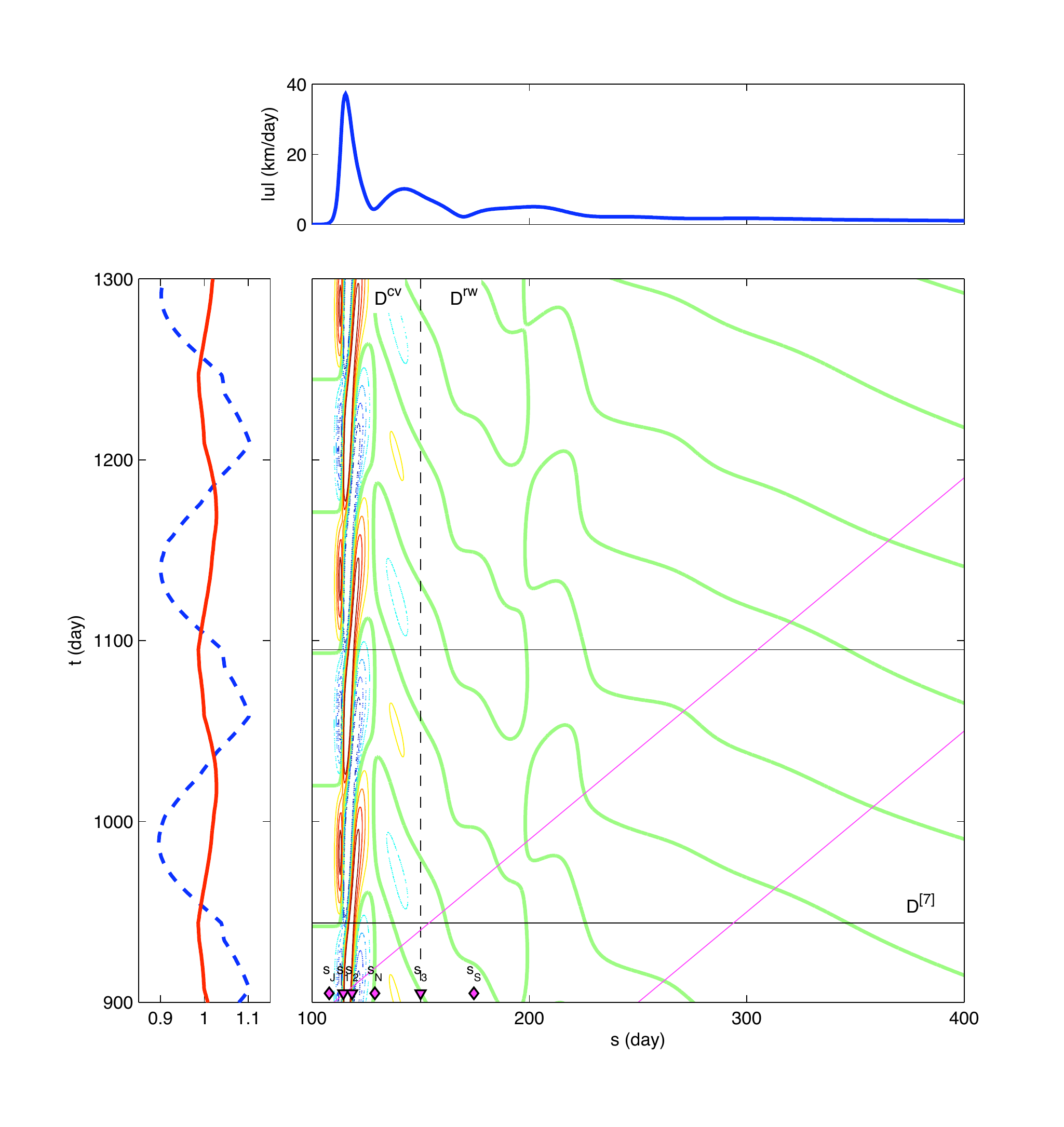}
\end{center}
\caption{The flux diagram $\mu^{C}(s,t)$ of the intergyre transport
 along the mean jet axis $C$
as the Hovm\"{o}ller diagram with $t$ runs vertically.
The contour interval is 20 (km$^2$ / day$^2$) with the dashed lines representing
 negative values. 
On the abscissa, the locations of 
$s_{{\rm J}=110}$, $s_{{\rm N}}=129$, and $s_{{\rm S}}=174.5$ are
 indicated by the diamonds, while
$s_{1}=114.5$, $s_{2}=118.5$,  and $s_{{\rm 3}}=150$ are indicated by
 the triangles.
In $\mu^{C}(s,t)$, $D^{[7]}$ are indicated by the two horizontal lines
 (solid) at
$t=t^{*}_{944}$ and $t=t^{*}_{944}(=t^{*}_{1095})$, 
while the boundary between
$D^{{\rm cv}}$ and $D^{{\rm rw}}$ is shown by the vertical line
 (dashed) at $s=s_{3}$.
The two diagonal lines (solid)
are examples of the reference trajectory, i.e.,
% (i.e., reference  particle advection), i.e.,
$(s_{0}-t_{0}+t,t)$  with $(s_{0},t_{0})=(s_{{\rm J}},900)$ and
$(s_{0},t_{0})=(250,900)$.
The left panel shows  $NT(t)$ (solid line) and
$TD(t)$ (dashed line) vs.~$t$.
The top panel shows 
$|\overline{{\bf u}}(\overline{{\bf x}}^{C}(s))|$ vs.~$s$ with the
same abscissa as  $\mu^{C}(s,t)$.}
\label{fg:dg_must_u}
\end{figure}

The Hovm\"{o}ller diagram 
\cite{hovmoller_tellus49,martis_etal_tellus06}
of the instantaneous flux in the $(s,t)$ space:
\begin{eqnarray} 
\mu^{C}(s,t) &\equiv& \phi(\overline{{\bf x}}^{C}(s),t) =
\overline{{\bf u}}(\overline{{\bf x}}^{C}(s))
\wedge{\bf u}'(\overline{{\bf x}}^{C}(s),t)~
\label{eq:mu_u} % \\
\end{eqnarray}
is fundamental to the TIME method because
it contains the stirring information locally and instantaneously 
extracted from $\phi({\bf x},t)$ along $C$. 
We refer to it as  the \emph{flux diagram}.
At a given instance $t$, 
a continuous segment of $s$ with $\mu^{C}(s,t)>0$
corresponds to a positive flux zone where
the instantaneous flux goes from the right to 
the left across $C$ with respect to the direction of increasing $s$. 
The direction of the flux is reversed for
$\mu^{C}(s,t)<0$.
The reference trajectory $(s,t)=(s_{0}-t_{0}+t,t)$
is a diagonal line going through 
$(s_{0},t_{0})$ with the unit slope
(see the main panel of Figure~\ref{fg:dg_must_u}).

The nature of the signals in $\mu^{C}(s,t)$ is dependent on both
the system and the choice of  $C$. 
These signals can be complex, as we shall observe in the double-gyre 
application (Figure~\ref{fg:dg_must_u}).
Nonetheless, having systematically
examined the mean $\overline{{\bf u}}({\bf x})$ (in terms of  
$\overline{\psi}({\bf x})$; Section~\ref{sec:bsc_mean}),
dynamic variability  ${\bf u}'({\bf x},t)$ (in terms of
 $\psi'({\bf x},t)$; Section~\ref{sec:bsc_dyn}),
flux variability $\phi({\bf x},t)$ (Section~\ref{sec:bsc_flx}), and
 the geographic  location of $C$ in the flow field 
(Section~\ref{sec:bsc_C}),  the physical
 interpretation of the  signals in $\mu^{C}(s,t)$ is straightforward.
Any signals  can be traced back to certain flux
zones
 in $\phi({\bf x},t)$,
 and hence the mean-eddy interaction process between 
$\overline{{\bf u}}({\bf x})$ and ${\bf u}'({\bf x},t)$.
%\fginsrt{fg:dg_must_u}

By the construction of the flux diagram along $C$,
the propagation speed and direction of the signals in $\mu^{C}(s,t)$
are defined with respect to the particle advection along the reference
trajectory of the mean flow. 
Signals with positive slopes correspond to the
downstream propagation of the coherent structures in $\psi'({\bf x},t)$.
In contrast, signals with negative slopes 
are related to the upstream propagation of the coherent structures. 
If the slope is steeper than 1, then the propagation speed is slower 
than the particle advection along $C$ in $\overline{\psi}({\bf x})$.

In the double-gyre application, $\mu^{C}(s,t)$ is 
almost periodic with period $T$ (Figure~\ref{fg:dg_must_u}), 
i.e., $\mu^{C}(s,t)=\mu^{C}(s,t+T)$, due to the
periodic dynamics in ${\psi}'({\bf x},t)$.
Within one period, the positive and negative phases are almost 
anti-symmetric, i.e., $\mu^{C}(s,t)\approx-\mu^{C}(s,t+T/2)$.
For $s<s_{{\rm J}}$, $\mu^{C}(s,t)$ is very small because of near-zero
$\overline{{\bf u}}(\overline{{\bf x}}^{C}(s))$ and changes 
the sign in synchrony with 
$T^{[k.1]}$ for $\mu^{C}(s,t)>0$ and $T^{[k.2]}$ for  $\mu^{C}(s,t)<0$.

The strongest signals in $\mu^{C}(s,t)$ are concentrated over
the segment $[s_{{\rm J}},s_{{\rm N}}]$ 
where the mean jet in $\overline{\psi}({\bf x})$ is fast
and the circulating eddy vortices in  ${\psi}'({\bf x},t)$ 
are energetic.
Due to the pulsation of the circulating eddy vortices as they propagate
along $C$ (Section~\ref{sec:bsc_flx}),
the corresponding flux zones also pulsate simultaneously 
over the three consecutive segments, 
$S_{a}\equiv [s_{{\rm J}},s_{1})$, 
$S_{b}\equiv [s_{1},s_{2})$, and 
$S_{c}\equiv [s_{2},s_{{\rm N}})$, with the alternating signs of
$\mu^{C}(s,t)$.
The widths of $S_{a}$, $S_{b}$, and $S_{c}$ are narrow
(6.5, 4, and 10.5days, respectively) with respect to the period of
intensification $T/2$ (75.5days)
during which the flux zones hardly move.
Thus the slopes of the dominant signals in $\mu^{C}(s,t)$ are steep.
At the end of $T^{[k.1]}$ and  $T^{[k.2]}$,
these flux zones weaken and quickly propagate downstream 
to the next segment along $C$.
A  positive flux zone over $S_{a}$ during $T^{[k.1]}$ connects to
$S_{b}$ during $T^{[k.2]}$, and then to $S_{c}$ during $T^{[k+1.1]}$
in a sequence;
conversely a negative flux zone over $S_{a}$ during $T^{[k.2]}$ 
moves over to $S_{b}$ during $T^{[k+1.1]}$, 
and then to $S_{c}$ during $T^{[k+1.2]}$.
%Each segment changes the sign of $\mu^{C}(s,t)$ every $T/2$.
Over the subsequent segment $S_{d}\equiv [s_{{\rm N}},s_{3})$ along $C$,
weak signals of the circulating eddy vortices
are observed in $\mu^{C}(s,t)$.
The slope is negative over $S_{d}$ because of the upstream propagation 
of the circulating eddies   as their centers leave $C$  
around ${\bf x}_{{\rm N}}$  (Figure~\ref{fg:dg_ph4_psie_phi}).

The westward propagation of the Rossby waves is
seen for  $s>s_{3}$.
The slope is negative and less than 1 because the Rossby waves 
propagate against the mean flow with the propagation speed 
faster than the particle advection along $C$. 
Magnitude of $\mu^{C}(s,t)$ rapidly decays to zero
for $s>500$ because of extremely small
$|\overline{{\bf u}}(\overline{{\bf x}}^{C}(s))|$ and
$|{\bf u}'(\overline{{\bf x}}^{C}(s),t)|$ there (not shown).
The change in the propagation slope over 
$s\in[200{\rm day},240{\rm day}]$ is mainly caused by the
meander of $C$
(Figure~\ref{fg:dg_me}).
Over $S_{e}\equiv[s_{{\rm S}},s_{N})$, 
the signals of the Rossby waves and those
of the circulating eddy vortices are mixed because the two types of
variability  merge 
between ${\bf x}_{{\rm N}}$ and  ${\bf x}_{{\rm S}}$ 
in ${\psi}'({\bf x},t)$
(Sections~\ref{sec:bsc_dyn} and \ref{sec:bsc_flx}).
% and $s_{3}$ in $[{\bf x}_{{\rm N}},{\bf x}_{{\rm S}}]$ is chosen  
% as the boundary point on $C$.

Accordingly, the $(s,t)$ space can be divided into sub-domains
 based on the types of the mean-eddy interaction. 
Table~\ref{tbl:D} summarizes the main domains for the
double-gyre application.
Two types of the variability are associated with the two domains;
 $D^{{\rm cv}}$ where the signals
are associated with the circulating eddy vortices in the dipole region
and $D^{{\rm rw}}$ where the signals are associated with the
Rossby wave in the downstream region of the eastward jet. 
The total domain is $D^{{\rm all}}=D^{{\rm cv}}\cup D^{{\rm rw}}$. 
Temporally periodic variability leads to the temporal decomposition of
$D^{{\rm all}}$ based on $T^{[k]}$,
i.e.,   $D^{{\rm all}}=\cup D^{[k]}$.
% where $D^{[k]}$ is the spatio-temporal  extension to $T^{[k]}$.
Another useful definition $D$ is transient $D(t)$ that 
covers the $(s,t)$ space up to the present time $t$.
\tblinsrt{tbl:D}

%For the variability associated with the circulating eddy vortices,
%we also define  eight simplified sub-domains 
%based on the flux zones of the circulating eddy vortices 
%($S_{a}$, $S_{b}$, $S_{c}$, and $S_{d}$)
%and two halves of the oscillation period
%($T^{[k.1]}$  and $T^{[k.2]}$).
%This simple division in $D^{[k]}$ is reasonable for describing the
%accumulation processes because these flux zones pulsate simultaneously and
%propagate much slower than the particle advection along $C$ in the mean
%flow.

% section 3: TIME function
\section{The TIME functions and the graphical approach to the analysis of
 transport processes}
\label{sec:fnc} 

The companion paper \cite{ide_wiggins_pd06a}
introduced the TIME functions. The main focus in 
 \cite{ide_wiggins_pd06a}  was the mathematical formulation, the validity of perturbation approximations, 
as well as the verification of the method by comparing with the
Lagrangian method based on the lobe dynamics for the case of $C$ chosen as a
heteroclinic connection of the reference state.
This section refines the TIME functions for a much more detailed
analysis of the transport processes due to the variability in the flow. 
We note that although the TIME functions as developed in \cite{ide_wiggins_pd06a} are based on a perturbation approach in the sense that the fluctuating part of the velocity field is ''small'' compared to the reference state, none of the development up to this point in the paper requires this smallness requirement--all that has been required is the decomposition of the velocity field into a (steady) reference state and a fluctuation about the reference state.

\subsection{The accumulation function}
\label{sec:fnc_m}

The accumulation function along $C$ is defined as
\begin{eqnarray}\label{eq:mC_org}
m^{C}(s,t;t_{0}:t_{1})&=&
\int_{t_{0}}^{t_{1}}\mu^{C}(s-t+\tau,\tau)~d\tau~.
\end{eqnarray}
The left-hand side of  (\ref{eq:mC_org}) denotes the amount of fluid
transport that occurs during the accumulation time interval
$[t_{0},t_{1}]$ evaluated at $(s,t)$. 
The right-hand side of (\ref{eq:mC_org}) expresses the transport
in terms of the spatio-temporal integration of
$\mu^{C}(s-t+\tau,\tau)$  during $t_{0}\leq \tau\leq t_{1}$.
Thus it can be thought that accumulation is advected, while it may
continue to occur, 
with the reference trajectory going through $(s,t)$.
The sign of
$m^{C}(s,t;t_{0}:t_{1})$ corresponds to the direction of transport
across $C$; $m^{C}(s,t;t_{0}:t_{1})>0$ is from right to left across $C$;
the direction of the flux is reversed for
$m^{C}(s,t;t_{0}:t_{1})>0$.

The basic idea of (\ref{eq:mC_org}) is that the transport 
can be estimated just for the time period of interest $[t_{0},t_{1}]$.
It can be extended to study ``when'', ``where'', and how'' variability of the
flow contributes to the transport.
This is done by restricting the integration in  (\ref{eq:mC_org})
to a specific space-time domain $D$
that contains the particular signals of interest 
(Section~\ref{sec:bsc_mu}).
Formally and practically this leads to the slight modification to the
accumulation function:
 \begin{subequations} \label{eq:maC} 
\begin{eqnarray}
m^{C}(s,t;D)
 =\int H(s-t+\tau,\tau;D) \mu^{C}(s-t+\tau,\tau) d\tau~,
 \label{eq:maC_d_m}
\end{eqnarray}
where
\begin{eqnarray}\label{eq:maC_d_H}
H(s-t+\tau,\tau;D)=\left\{\begin{array}{ll}
1 & \mbox{ if $(s-t+\tau,\tau)\in D$}, \\
0 & \mbox{ otherwise}~ \end{array} \right. 
\end{eqnarray} 
\end{subequations} 
acts as a switch to turn on and off the instantaneous flux depending on 
whether or not $(s-t+\tau,\tau)$ is in $D$ at time $\tau$.
For example using $D(t)$ as $D$, $m^{C}(s,t;D(t))$ is the transient 
transport by accounting for the accumulation up to the present time $t$
(see Table~\ref{tbl:D} for the definition of $D(t)$).

For the analysis of transport processes, 
there are properties of $m^{C}(s,t;D)$  that are useful
(see \cite{ide_wiggins_pd06a} for technical details). 
One is the \emph{invariance} property along the 
individual reference trajectory by advection of the accumulation:
\begin{eqnarray}\label{eq:mCdlt}
m^{C}(s,t;D)&=&m^{C}(s+\delta,t+\delta;D) 
\end{eqnarray} 
for any $\delta$, but with $D$ fixed. 
The invariance property implies a conservation law of the accumulation
by the advection.  At any $(s,t)$ along a reference trajectory, 
$m^{C}(s,t;D)$ is independent of $t$ as long as $D$ is independent of $t$.

The other useful property is the  \emph{(piece-wise) independence} property
of $D$. By breaking up the domain $D$ into $L$ non-overlapping
sub-domains $D_{1}, \ldots,D_{L}$, the piece-wise independence
property implies that the accumulation function can be written as

\begin{eqnarray}\label{eq:mCk} 
m^{C}(s,t;D)&=&\sum_{l=1}^{L}m^{C}(s,t;D^{l})~.
\end{eqnarray} 

\noindent
For example, in the double-gyre application, the effect of the
circulating eddy vortices and that of the Rossby wave propagation can
be examined separately, while  the overall transport is the sum of
the two, i.e., 
$m^{C}(s,t;D^{{\rm all}})=m^{C}(s,t;D^{{\rm cv}})+m^{C}(s,t;D^{{\rm rw}})$.
Transient transport can be also decomposed into the sub-domains
 as long as $t$ is the same for all,
e.g., $m^{C}(s,t;D^{{\rm all}}(t))=\sum_{k}m^{C}(s,t;D^{[k]}(t))$.

A key feature of the TIME method is the 
\emph{graphical approach}
using the Hovm\"{o}ller diagram of $\mu^{C}(s,t)$. 
By  definition (\ref{eq:maC}),
transport processes are described by the manner in which 
the reference trajectory passes through the 
signals in $\mu^{C}(s,t)$ over the specific domain of interest $D$
(see Figure\ref{fg:dg_must_u}).
Conversely, it discloses the dynamical origins of $m^{C}(s,t;D)$ by
relating the signals in $\mu^{C}(s,t)$ to dynamic activities in
${\bf u}'({\bf x},t)$ through $\phi({\bf x},t)$
(Section~\ref{sec:bsc}). Thus,
the graphical approach unveils the dynamical processes of the
mean-eddy interaction that are responsible for transport.

Moreover, the graphical approach can assess the efficiency of the
dynamic activity ${\bf u}'({\bf x},t)$ in contributing to $m^{C}(s,t;D)$.
The accumulation is, in general, extremely efficient if a signal in
$\mu^{C}(s,t)$ propagates with the same unit slope
because it keeps accumulating the same signed instantaneous flux.
In other words, transport processes are the most efficient if 
the (Eulerian) eddy activity in ${\psi}'({\bf x},t)$  
propagates  at the same speed as the (Lagrangian) particle 
advection of the mean  flow  $\overline{{\psi}}({\bf x})$ along $C$.
In contrast, if the eddy activity is temporally 
periodic and propagates upstream of the mean
flow, then the net contribution  to the accumulation adds up to
zero. This is because the reference trajectory
cuts across the signals of $\mu^{C}(s,t)$ whose sign alternates
periodically, resulting in the cancellation of the accumulation.

\subsection{The displacement function}
\label{sec:fnc_r}

The second type of  TIME function quantifies  the geometry
associated with transport, up to the leading-order in the ``size''
of the unsteady part of the velocity field. To describe the
geometry in the two-dimensional flow, we use an orthogonal
coordinate set $(l,r)$ for ${\bf x}$ near $C$,
where the arc-length coordinate $l=l^{C}(s)$ of 
$\overline{{\bf x}}^{C}(s)$ is defined along  $C$ by 
$\frac{d}{ds}l^{C}(s)=|\overline{{\bf u}}(\overline{{\bf x}}^{C}(s))|$.
By taking the normal projection $\overline{{\bf x}}^{C}(s)$ of 
${\bf x}$ onto $C$,
the signed distance coordinate $r=r^{C}(l^{C}(s))$ is defined by
$r^{C}(l^{C}(s))\equiv(\overline{{\bf u}}(\overline{{\bf x}}^{C}(s))/
|\overline{{\bf u}}(\overline{{\bf x}}^{C}(s))|)
\wedge({\bf x}-\overline{{\bf x}}^{C}(s))$,
where $|\overline{{\bf u}}(\overline{{\bf x}}^{C}(s))|\neq 0$. 
If $r^{C}(l^{C}(s))>0$, then ${\bf x}$ lies in the left side of $C$ 
with respect to the direction
defined by the direction of increasing $s$. The side of ${\bf x}$ is
reversed for $r^{C}(l^{C}(s))<0$.

The displacement distance function is defined as 

\begin{subequations}\label{eq:ra}
\begin{eqnarray}\label{eq:ra_d} 
r^{C}(l^{C}(s),t;D)
 &=&\frac{a^{C}(s,t;D)}{|\overline{{\bf u}}(\overline{{\bf x}}^{C}(s))|}~,
\end{eqnarray}
where
\begin{eqnarray}
a^{C}(s,t;D)&=& \int
H(s-t+\tau,\tau;D)
\overline{ e}^{C}(s-t+\tau:s) \mu^{C}(s-t+\tau,\tau)
d\tau~:
\label{eq:ra_a}
\end{eqnarray}

\noindent
is the displacement area per unit $s$
and $H$ is defined in (\ref{eq:maC_d_H}).
Compressibility of the flow is taken into account by

\begin{eqnarray}\label{eq:ra_ae}
\overline{ e}^{C}(s-t+\tau:s)&=&
\exp\left\{\int_{s-t+\tau}^{s}
D_{\bf x}{\bf u}
(\overline{{\bf x}}^{C}(\sigma))d\sigma\right\}~,
\end{eqnarray}
\end{subequations}

\noindent
so that the flux that has occurred at $(s-t+\tau,\tau)$ may
expand or compress as it advects to $(s,t)$ with the reference particle
advection. 
If the mean flow is incompressible as in the quasi-geostrophic model
used for the double-gyre application, then $m^{C}(s,t;D)$ and
$a^{C}(s,t;D)$ are the same since $\overline{ e}^{C}(s-t+\tau:s)\equiv
1$ holds for any $s-t+\tau$ and $s$. The technical details can
be found in \cite{ide_wiggins_pd06a}.

The advantage of $r^{C}(l,t;D)$ is that  
it gives the geometry of the so-called ``pseudo-lobes'' that represent 
the spatial coherency of the transport. Pseudo-lobes are the
chain-like geometry of transport defined by the areas surrounded by
$C=\{(l,r)~|~r=0\}$ and the curve $\{(l,r)~|~r=r^{C}(l,t;D)\}$
 at a given time $t$.
A positive pseudo-lobe corresponds to a coherent area defined by
$\{(l,r)~|~0\leq r \leq r^{C}(l,t;D)\}$, while  a negative
pseudo-lobe is a coherent area defined by
$\{(l,r)~|~r^{C}(l,t;D)\leq r \leq 0\}$. 
The relation of the pseudo-lobes to the Lagrangian lobes is discussed
in \cite{ide_wiggins_pd06a}.

% section 4: DG application
\section{Application to the Inter-gyre Transport in the Double-Gyre Ocean}
\label{sec:dg}

Having carried out the preparatory steps (Section~\ref{sec:bsc})
 and defined the TIME functions (Section~\ref{sec:fnc}), 
we now use the TIME method to analyze  
inter-gyre transport processes in the the double-gyre circulation. 
We focus on the following three aspects of transport processes
using the  specific types of  sub-domain $D$ for $m^{C}(s,t;D)$
(see  Table~\ref{tbl:D} for the definition of these sub-domains): 
using $D(t)$, we study how the variability give rise to the development
 of the pseudo-lobes;
using $D^{{\rm cv}}(t)$ and  $D^{{\rm rw}}(t)$,
we compare the impact of the circulating eddy vortices and 
that of the Rossby waves  to the inter-gyre transport;
using $D^{{\rm all }}$ along with  $D(t)$ and $D^{[k]}$, 
we analyze the inter-gyre transport processes of the Lagrangian lobes
and examine the impact of  variability during $T^{[k]}$.
We also use $m^{C}(s,t;D)$ in place for
$a^{C}(l,t;D)$, with which we describe the geometry 
$r^{C}(l,t;D)$ associated
with the transport, particular in terms of the pseudo-lobes.

\begin{figure}[!ht]
\begin{center}
\includegraphics[height=10.5cm]{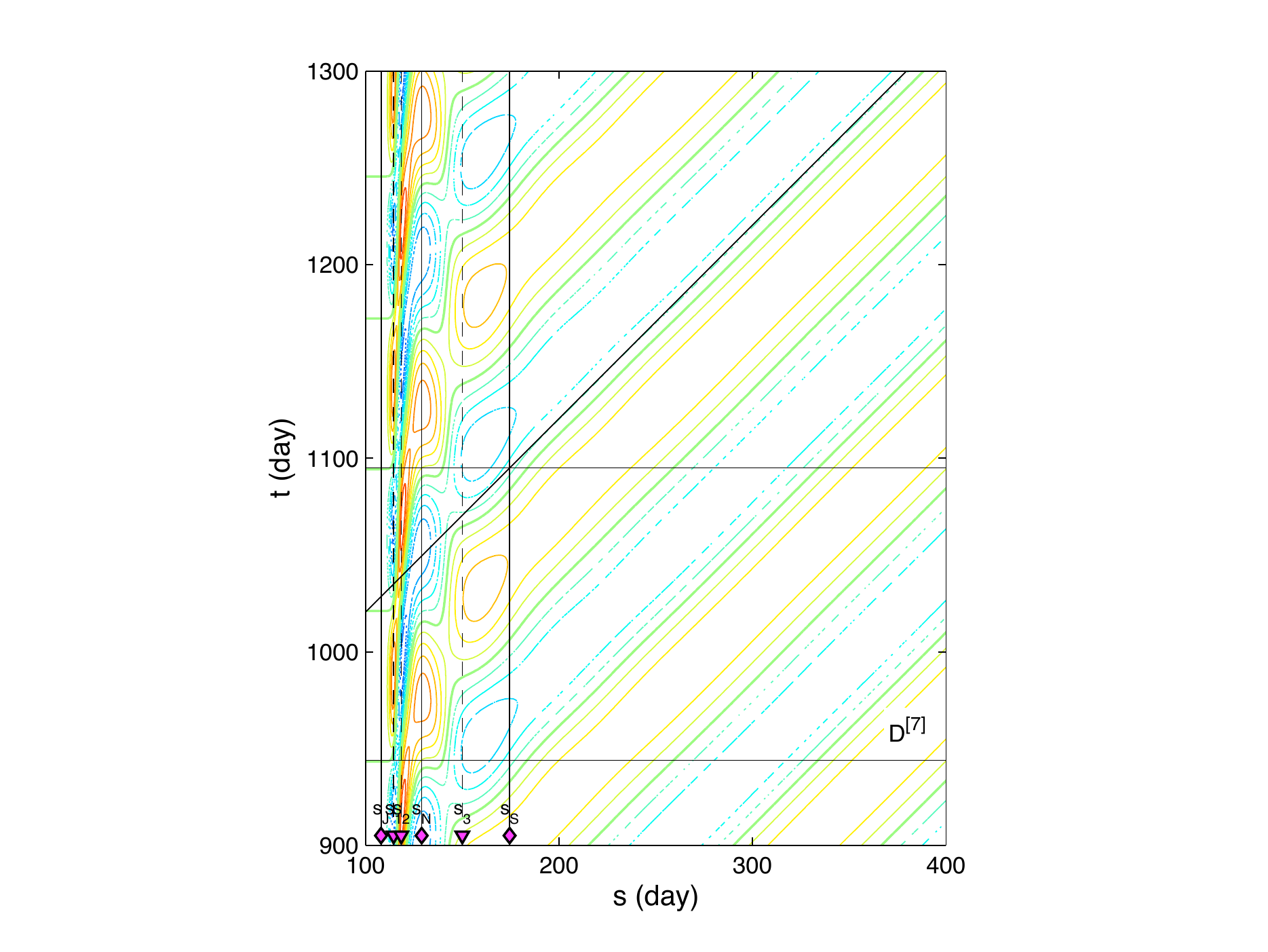}
\end{center} 
\caption{Transient accumulation $m^{C}(s,t;D(t))$.
The contour interval is 50 with the dashed contours for the negative
 values; the diamonds show
$s_{{\rm J}}$, $s_{{\rm N}}$, and  $s_{{\rm S}}$
while the triangles show
$s_{1}$, $s_{2}$, and  $s_{3}$
from upstream to downstream.
A reference trajectory is plotted for 
$(s,t)=(s_{0}-t_{0}+t,t)$ that goes through $s_{{\rm S}}$
at the end of $T^{[7]}$ with 
$(s_{0},t_{0})=(s_{{\rm  S}},t^{*}_{944}+T)$.}
\label{fg:dg_mDt}
\end{figure}

Figure~\ref{fg:dg_mDt} shows the transient accumulation
$m^{C}(s,t;D(t))$  in the Hovm\"{o}ller diagram
from the past up to the evaluation (present) time 
$t$ as $t$ progresses upward.
\fginsrt{fg:dg_mDt}
Due to the nearly periodic variability in ${\psi}'({\bf x},t)$, 
$m^{C}(s,t;D(t))$ is also nearly 
periodic in $t$, i.e., $m^{C}(s,t;D(t))=m^{C}(s,t+T;D(t+T))$. 
Because of the anti-symmetry in the oscillation,
$m^{C}(s,t;D(t))$ is also anti-symmetric with respect to a $T/2$-shift in $t$,
i.e., $m^{C}(s,t;D(t))=-m^{C}(s,t+T/2;D(t+T/2))$.
For $s<s_{{\rm J}}$ at any $t$,  $m^{C}(s,t;D(t))$ is nearly zero
because hardly any accumulation happens in the upstream direction of
$s_{{\rm J}}$ due to extremely small $\mu^{C}(s,t)$ there.
For $s>s_{{\rm S}}$,
$m^{C}(s,t;D(t))\approx  m^{C}(s+\delta,t+\delta;D(t+\delta))$
along the reference trajectory 
(e.g., diagonal line in Fig~\ref{fg:dg_mDt})
with $\delta>0$
indicates very little accumulation there.
It suggests that the Rossby waves contribute very little to
the inter-gyre transport in the downstream direction of $s_{{\rm S}}$.

\begin{figure}[!ht]
\begin{center}
\includegraphics[height=10.cm]{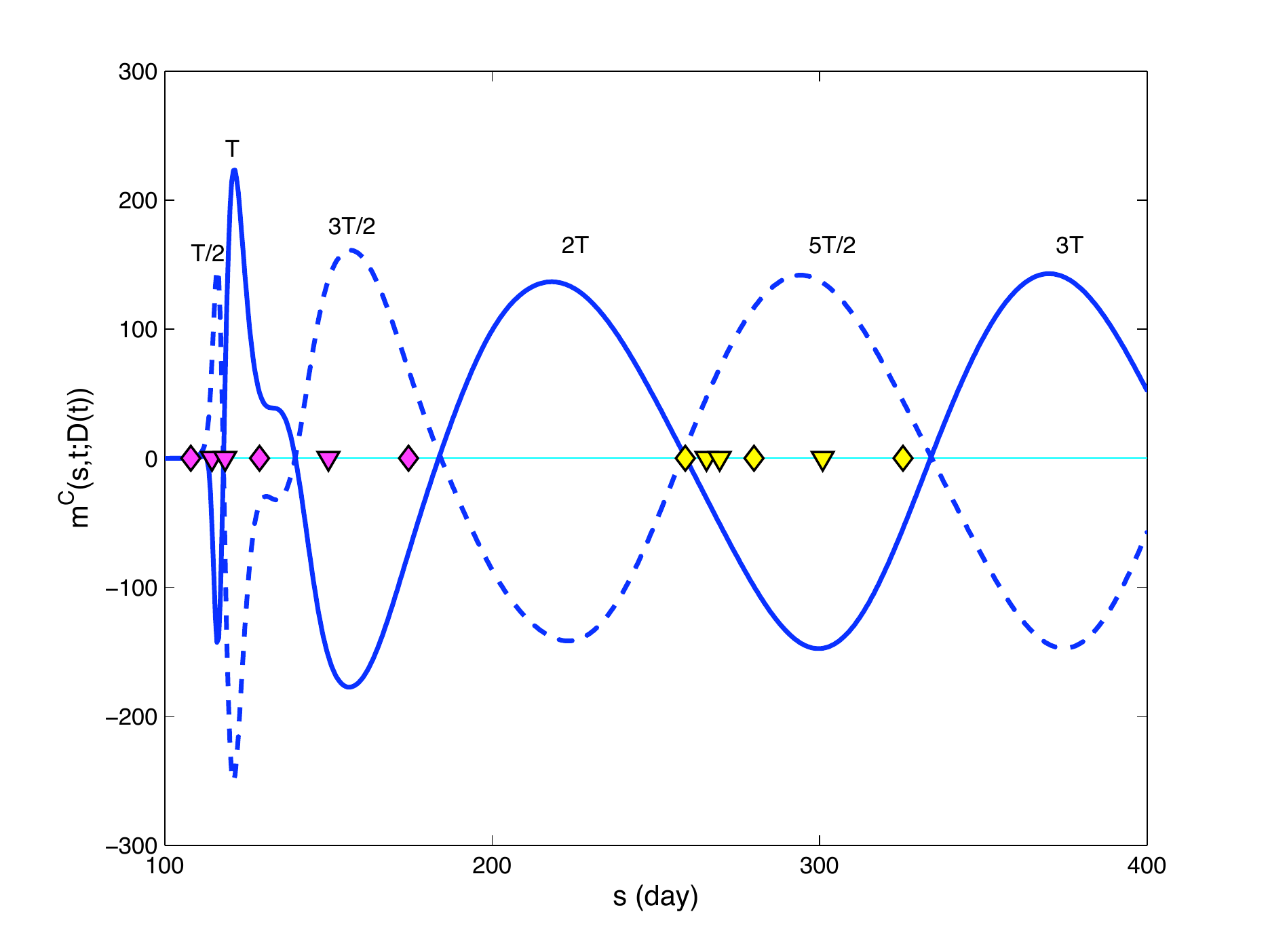}
\end{center} 
\caption{Pseudo-lobes of  $m^{C}(s,t;D(t))$
 at the end of $T^{[k.1]}$ ($t=t^{*}_{36}+(k-1/2)T$; dash line)
and at the end of $T^{[k.2]}$ ($t^{*}_{36}+kT$; solid line) 
for $k\geq 1$; the figure is made using $k=7$.
The positive pseudo-lobe that starts developing from
the begining of  $T^{[7.1]}$ 
at $t^{*}_{944}$($=t^{*}_{36}+(k-1)T$ with $k=7$) is indicated by: 
$T/2$ at $t=t^{*}_{944}+T/2$;
$T$ at $t=t^{*}_{944}+T$;   
$3T/2$ at $t^{*}_{944}+3T/2$;  
and $2T$ at $t^{*}_{944}+2T$.
Dark diamonds show
$s_{{\rm J}}$, $s_{{\rm N}}$, and  $s_{{\rm S}}$;
dark triangles show
$s_{1}$, $s_{2}$, and  $s_{3}$;
lighter diamonds show
$s_{{\rm J}}+T$, $s_{{\rm N}}+T$, and  $s_{{\rm S}}+T$;
and light triangles show
$s_{1}+T$, $s_{2}+T$, and  $s_{3}+T$.}
\label{fg:dg_mDt_PL}
\end{figure}

By a comparison of Figure~\ref{fg:dg_mDt} with
Figure~\ref{fg:dg_must_u}, accumulation processes in
$m^{C}(s,t;D(t))$ are closely related to the
evolution of the flux zones  in $\mu^{C}(s,t)$ 
(Section~\ref{sec:bsc_mu}).
Below, we follow the development of a positive pseudo-lobe 
in  Figure~\ref{fg:dg_mDt}
starting from $T^{[7.1]}$ as $t$ progresses.
This pseudo-lobe first gains positive accumulation 
over $S_{{\rm a}}$ where a positive flux zone pulsates 
during $T^{[7.1]}$.
If there is no flux in the downstream of $S_{{\rm a}}$ 
(i.e., $\mu^{C}(s,t)=0$  for $s>s_{1}$), 
then this positive pseudo-lobe will spread over 
$[s_{{\rm J}},s_{1}+T/2]$ at the end of $T^{[7.1]}$
because the accumulation is simply advected along
the reference trajectory; 
this corresponds to the invariance property (\ref{eq:mCdlt}).
However, positive accumulation that occurs over 
$S_{{\rm a}}$ is canceled shortly after it is advected into
$S_{{\rm b}}$ where a strong negative flux zone pulsates during $T^{[7.1]}$.
Hence,  a small, narrow (in terms of $s$) positive pseudo-lobe develops
mostly over $S_{{\rm a}}$ during $T^{[7.1]}$.
At the end of $T^{[7.1]}$ when the positive flux zone over $S_{{\rm a}}$ moves
quickly  to $S_{{\rm b}}$,
the positive pseudo-lobe follows it.
In Figure~\ref{fg:dg_mDt_PL}, 
$m^{C}(s,t;D(t))$ at $t=t^{*}_{944}+T/2$ is shown by the dashed line and 
this positive pseudo-lobe is indicated by $T/2$.
\fginsrt{fg:dg_mDt_PL}

During $T^{[7.2]}$,
the positive pseudo-lobe continues to develop 
over $S_{{\rm b}}$ where  the positive flux zone pulsates with large
amplitude.
Because this flux zone is stronger than any flux zones,
the pseudo-lobe slowly spreads into the downstream region by the
advection of the accumulation.
At the end of $T^{[7.2]}$ when the positive flux zone quickly moves 
to $S_{{\rm c}}$,
the positive pseudo-lobe once again follows it.
In Figure~\ref{fg:dg_mDt_PL}, $m^{C}(s,t;D(t))$
is plotted at $t=t^{*}_{944}+T$  by the solid line.
The spread of the pseudo-lobe  is observed over $S_{{\rm d}}$.

During $T^{[8.1]}$, the pseudo-lobe continues to develop 
over $S_{{\rm c}}$ where the positive flux zone 
pulsates before disappearing at $s_{{\rm N}}$. 
It also continues to spread into the downstream region
because the flux zones are much weaker there.
Speed of the spread increases in the downstream direction of $s_{3}$ 
because of a weaker, positive flux zone
induced by the Rossby waves  over $S_{{\rm e}}$.
At the same time, the spread of the negative pseudo-lobe that started in 
$T^{[7.2]}$ gradually pushes this positive pseudo-lobe from the upstream region.
At the end of $T^{[8.1]}$, 
the main part of the positive pseudo-lobe is located over 
$S_{e}$ (Figure~\ref{fg:dg_mDt_PL}).

During  $T^{[8.2]}$, the positive pseudo-lobe first gains 
accumulation from a positive flux zone induced by the Rossby waves
over $S_{{\rm e}}$, but loses it quickly 
because the negative flux zone induced by the Rossby waves moves into 
$S_{{\rm e}}$.
In the meantime, the positive pseudo-lobe continues to spread into the
downstream region. 
Once the accumulation passes $s_{{\rm S}}$, it simply advects along the
reference trajectory.
Near the end of $T^{[8.2]}$, the positive pseudo-lobe reaches the final
form as the entire pseudo-lobe passes  $s_{{\rm S}}$.

Our analysis of $m^{C}(s,t;D(t))$ therefore reveals that the
majority of  inter-gyre transport occurs over a very limited
segment $[s_{{\rm J}},s_{{\rm S}}]$.
Accumulation processes are synchronized with the evolution of the
circulating eddy vortices, and are fed mostly 
by the three flux zones over $3T/2$ in $t$ while they are mostly
the same signed over the spatial 
segments  $S_{{\rm a}}$ through $S_{{\rm c}}$;
in the subsequent $T/2$ in $t$, the Rossby waves help form the final shape of
the pseudo-lobe with the width $T/2$ in $s$.
After $2T$ in $t$, the pseudo-lobe moves by the advection
along the reference trajectory
because there are no further accumulation in the downstream direction
of $S_{{\rm S}}$.
Thus, it takes almost $2T$ in $t$
to develop a fully-grown pseudo-lobe of the width $T/2$ in $s$.
While developing, the propagation speed of the pseudo-lobe is 
much slower than the particle advection along the reference trajectory:
over $2T$ in $t$, the pseudo-lobe moves about $T/2$ in 
$s$ starting from $S_{{\rm a}}$.

\begin{figure}[!ht]
\begin{center}
\includegraphics[height=10.cm]{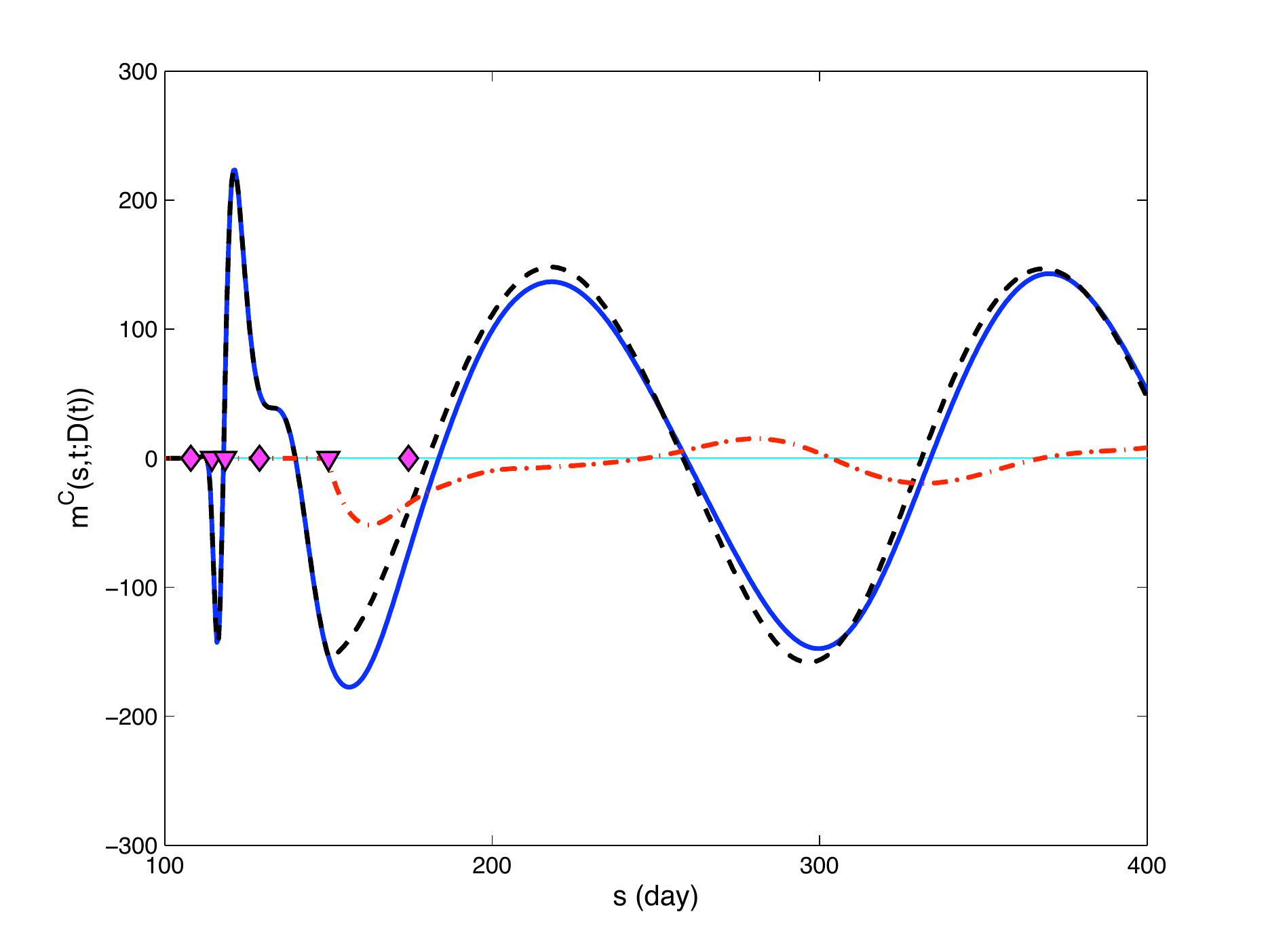}
\end{center}
\caption{Pseudo-lobes  of 
$m^{C}(s,t;D(t))$ for the total transient transport
 (solid line; same as in Figure~\ref{fg:dg_mDt_PL}),
 $m^{C}(s,t;D^{{\rm cv}}(t))$ 
 by the circulating eddy vortices (dash line),
 and $m^{C}(s,t;D^{{\rm rw}}(t))$ 
 by the circulating eddy vortices  (dash-dot line) 
 at $t=t^{*}_{36}+kT$ (i.e., the end of $T^{[k]}$);
 the figure is made at $t^{*}_{1095}$ using  $k=7$.
 Dark diamonds and dark 
triangles are the same as in Figure~\ref{fg:dg_mDt_PL}.}
\label{fg:dg_mDt_PL_allcvrw}
\end{figure}

The analysis also reveals
that the Rossby waves have very little impact on the inter-gyre
transport for two reasons. 
One reason is that the  signals of
$\mu^{C}(s,t)$ in $D^{{\rm rw}}$ are weak
 (see Section~\ref{sec:bsc_mu}). 
The other reason comes
from the fact that the upstream propagation of the signal in
$\mu^{C}(s,t)$ in  $D^{{\rm rw}}$ is  nearly periodic in time
(see Section~\ref{sec:fnc_m}).
Figure~\ref{fg:dg_mDt_PL_allcvrw} shows the decomposition of the
transient inter-gyre transport $m^{C}(s,t;D(t))$ into the
part induced by the circulating eddy vortices  
$m^{C}(s,t;D^{{\rm cv}}(t))$ and that 
induced by the Rossby waves
$m^{C}(s,t;D^{{\rm rw}}(t))$ at  $t=t^{*}_{36}+kT$ (end of $T^{[k]}$), 
where $m^{C}(s,t;D(t))$
 $=m^{C}(s,t;D^{{\rm cv}}(t))$  $+m^{C}(s,t;D^{{\rm rw}}(t))$
by the piece-wise independence property (\ref{eq:mCk}).
\fginsrt{fg:dg_mDt_PL_allcvrw} 
By the definition of $D^{{\rm rw}}$, 
$m^{C}(s,t;D^{{\rm rw}}(t))=0$ for $s<s_{3}$ means that
the Rossby waves impact the
inter-gyre transport only in the downstream of $s_{3}$
(dash-dot line).
The amount of  accumulation, $m^{C}(s,t;D^{{\rm rw}}(t))$,
is small except over $[s_{3}, s_{{\rm S}}]$ where the signals
of the circulating eddy vortices and  that of the Rossby waves are 
mixed in $\mu^{C}(s,t)$ (Section~\ref{sec:bsc_mu}).

\begin{figure}[!ht]
\begin{center}
\includegraphics[height=10.cm]{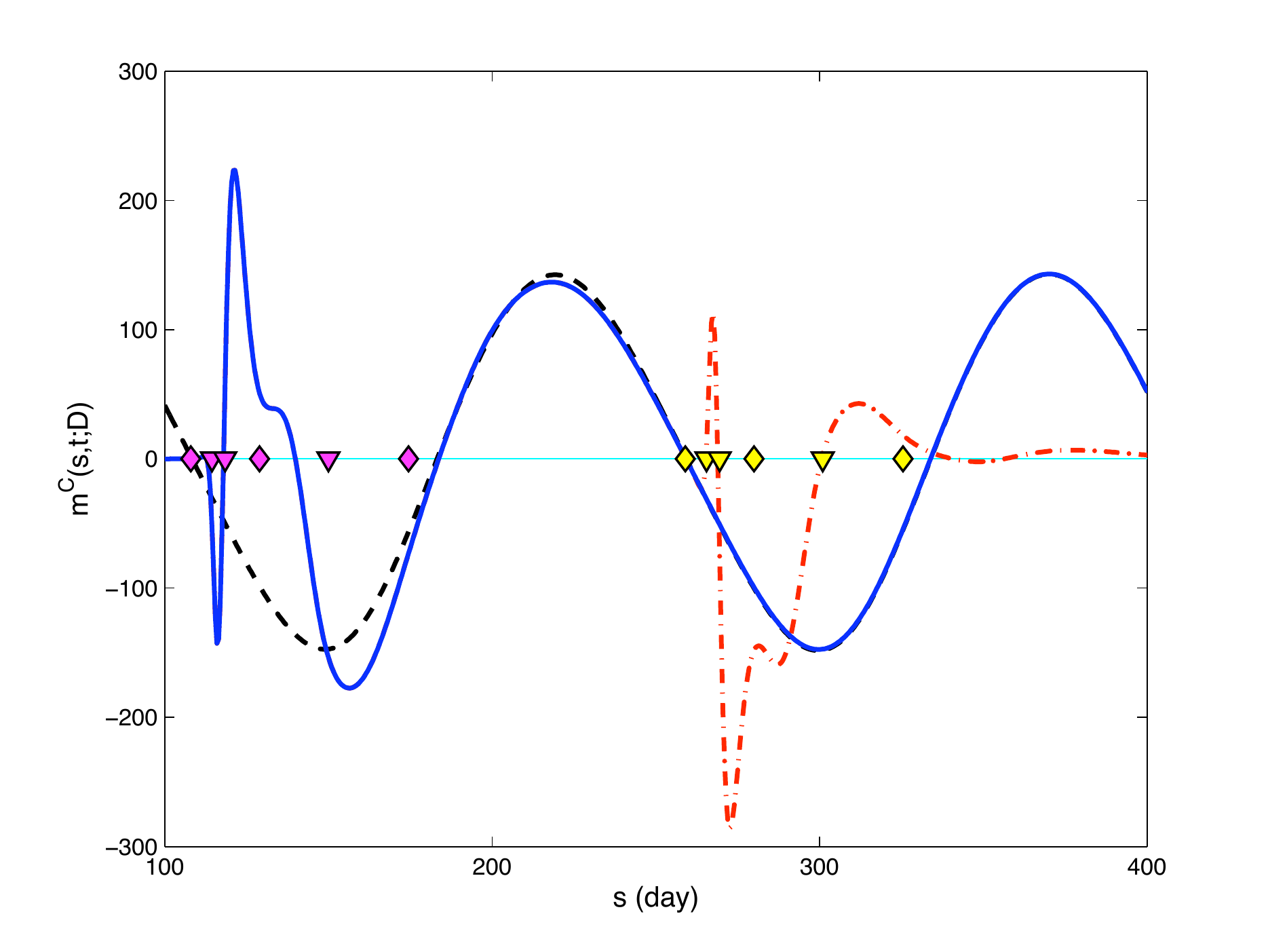}
\end{center} 
\caption{Pseudo-lobes  of  
$m^{C}(s,t;D(t))$ for the total transient transport
 (solid line; same as in Figures~\ref{fg:dg_mDt_PL}),
$m^{C}(s,t;D^{[k]})$ for transport induced during 
$T^{[k]}$ (dash-dot line),
$m^{C}(s,t;D^{{\rm all}})$  for total transport (dash line) 
 at $t=t^{*}_{36}+kT$ (i.e., the end of $T^{[k]}$);
 the figure is made at $t^{*}_{1095}$ using $k=7$.
Diamonds and triangles are the same as in Figure~\ref{fg:dg_mDt_PL}.}
\label{fg:dg_mD_PL}
\end{figure}

Lagrangian lobes have proven to be extremely useful and insightful in a variety of transport studies
and they provide precise amounts of  Lagrangian transport of fluid
\cite{malhotra_wiggins_jnls98,msw,samwig,wiggins_arfm05}.
In the companion paper \cite{ide_wiggins_pd06a}, it was shown that 
$m^{C}(s,t;D^{{\rm all}})$ provides a good approximation 
to the amount of transport carried by individual lobes in the inter-gyre
transport (see also \cite{coulliette_wiggins_npg00}).
Figure~\ref{fg:dg_mD_PL}  shows $m^{C}(s,t;D^{{\rm all}})$ by the dash
line at the end of $T^{[k]}$. 
It is doubly periodic in $s$ and $t$ because of 
the periodic dynamics in $\psi'({\bf x},t)$.

Using the TIME method, we analyze the transport processes associated
with the Lagrangian lobes.
The transient transport $m^{C}(s,t;D(t))$ in the past up to the present 
time $t$ is shown in Figure~\ref{fg:dg_mD_PL} by the solid line.
By the piece-wise independence (\ref{eq:mCk}),
the  difference  $m^{C}(s,t;D^{{\rm all}})-m^{C}(s,t;D(t))$
corresponds to the transport that will occur in the future of $t$.
A significant difference is observed mainly for $s<s_{{\rm N}}$
where $m^{C}(s,t;D(t))$ does not include the active accumulation
 over $[s_{{\rm J}},s_{{\rm N}}]$ yet.
A slight difference  occurs 
over $[s_{{\rm N}},s_{{\rm J}}+T]$  due to the flux zones of the
Rossby waves there.
For $s>s_{{\rm J}}+T$,
$m^{C}(s,t;D^{{\rm all}})$ and  $m^{C}(s,t;D(t))$ are almost
indistinguishable, suggesting that no further transport will occur in
the future over the segment of $C$.
Once again we confirm that the impact of Rossby waves
in the downstream region is negligible for  inter-gyre transport.
Note that 
 $m^{C}(s,t;D^{{\rm cv}})\approx m^{C}(s,t;D^{{\rm cv}}(t))$  and
 $m^{C}(s,t;D^{{\rm rw}})\approx m^{C}(s,t;D^{{\rm rw}}(t))$ hold
 over there as well (see Figure\ref{fg:dg_mDt_PL_allcvrw}).
\fginsrt{fg:dg_mD_PL} 

To examine how much transport occurs during one period 
$T^{[k]}$ of the unsteady eddy activity in $\psi'({\bf x},t)$,
Figure~\ref{fg:dg_mD_PL} shows $m^{C}(s,t;D^{[k]})$ by the dash-dot line.
Over $[s_{{\rm J}},s_{{\rm J}}+T]$ of the length $T$ in $s$,
$m^{C}(s,t;D^{[k]})$ and $m^{C}(s,t;D(t))$ are almost
indistinguishable because both include the active accumulation 
 over $[s_{{\rm J}},s_{{\rm N}}]$.
Accordingly,  during $T^{[k]}$, only one positive pseudo-lobe grows 
close to its final form over $[s_{{\rm J}}+T/2,s_{{\rm J}}+T]$ with
width $T/2$ in $s$, but no negative pseudo-lobe can grow to the final
form. If the period is shifted by $T/2$, then only one negative
pseudo-lobe grows into its final form over the same 
$[s_{{\rm J}}+T/2,s_{{\rm J}}+T]$ segment.
For $s>s_{{\rm J}}+T$, $m^{C}(s,t;D^{[k]})$ differs from 
$m^{C}(s,t;D(t))$ because  of the accumulation in 
$m^{C}(s,t;D(t))$  that has occurred prior to $T^{[k]}$.

% section 5: conclusion
\section{Concluding remarks}
\label{sec:cncl}

Building on the transport method developed in the companion paper
 \cite{ide_wiggins_pd06a},
 we have formulated a framework for the analysis of  the dynamical 
processes that influence transport.
 The transport method, called the 
``{\bf T}ransport {\bf I}nduced by the {\bf M}ean-{\bf E}ddy
 {\bf i}nteraction'' (TIME), is a hybrid combination of Lagrangian and Eulerian
 transport approaches.
Our analysis proceeds by a step by step approach. In particular, 
the steps are to determine the mean flow structure of
$\overline{{\bf u}}({\bf x})$,
 determine the dynamic variability in ${\bf u}'({\bf x},t)$,
 construct the instantaneous stirring chart 
$\phi({\bf x},t)=\overline{{\bf u}}({\bf x})\wedge{\bf u}'({\bf x},t)$
 induced by the mean-eddy interaction,
 choose an Eulerian boundary $C=\{\overline{{\bf x}}^{C}(s)\}$, 
and compute the
 flux diagram $\mu^{C}(s,t)=\phi(\overline{{\bf x}}^{C}(s),t)$.
The signals in $\mu^{C}(s,t)$ are define relative to the reference
particle advection along $C$, which is a diagonal line in the
Hovm\"{o}ller diagram.

The fundamental constructions underlying the TIME method involve
computing  transport, either as accumulation
 $m^{C}(s,t;D)$ or displacement area $a^{C}(l,t;D)$ 
(which gives displacement
 distance $r^{C}(l,t;D)$),  which rely on the
 spatio-temporal integration of $\mu^{C}(s,t)$.
This provides the two-way link
between the
 variability of the flow and  the actual  transport processes.
It is a unique feature of the TIME method that neither the
Eulerian nor
 Lagrangian methods alone  can provide.
These fundamentals also provides  a platform for a novel graphical
approach to the analysis of transport processes.

While transport is highly system dependent, there are some common
features that can hold in general that we can understand from our
graphical approach for the analysis of transport processes. For
example, the accumulation is most effective if the signals in
$\mu^{C}(s,t)$ propagate with the same unit slope as  the reference
trajectory, i.e., dynamic variability propagates with the particle advection in the
mean flow. 
However, if the dynamic variability is temporally near
periodic, and the signal has a negative slope, then the net effect
would be almost zero. This may happen when the wave propagates
upstream in the mean flow, like in the westward Rossby wave
propagation in the double-gyre application. The role of
variability in transport is analytically studied in
\cite{ide_wiggins_prep06} and is based on a related  spatio-temporal scale analysis.

 We have applied our  framework to the analysis of intergyre transport
 processes in  the double-gyre ocean circulation where
 the Rossby-wave mode dominates the dynamic variability with a
 period $T$.
 The spatio-temporal analysis shows that the intergyre transport is
 controlled by a complex rotation of eddy vortices in the fast
 eastward jet near the western boundary current.
 The emergence of the pseudo-lobes is synchronized
 with the circulating eddy vortices in ${\bf u}'({\bf x},t)$.
 Pseudo-lobes having  alternating signed area emerge every $T/2$ to transport
 water across the mean jet axis between the subpolar and subtropical
 gyres, while  each pseudo-lobe sepends almost $2T$
 over a very limited segment in the upstream dipole region
 to fully develop. During the development period, the pseudo-lobes
 propagate at a much slower speed than the reference particle
 advection.
 However, once fully developed, they propagate downstream of the mean
 jet axis at the same speed as the reference particle advection.
 The basin-scale Rossby wave has very little impact on the intergyre
 transport.

% acknowledgement
%\addcontentsline{toc}{section}{\protect\numberline{}{Acknowledgment}}
%\addcontentsline{toc}{section}{\protect\numberline{}{Acknowledgment}}
\section*{Acknowledgment}
This research is supported by ONR Grant No.~ N00014-09-1-0418, (KI)
and ONR Grant No.~N00014-01-1-0769 (SW).

% \appendix
%  Appendix
%\clearpage
%\appendix
%\input{time_II_ar}

\newpage
\bibliographystyle{elsart-num}

\newpage
\addcontentsline{toc}{section}{\protect\numberline{}{Tables}}
\listoftables
\clearpage
%% TABLES

%% table: D - - - - - - - - - - - - - - - - - - -
\begin{table}[htb!]
\begin{center}
\begin{tabular}{|c|p{3in}l|l|} \hline
Domain & Description & Definition \\ \hline
$D^{{\rm cv}}$ & domain associated with the circulating eddy vortices
     & $\{(s,t)~|~ s< s_{{\rm tr}}\}$ \\ 
$D^{{\rm rw}}$ & domain asscoiated with the Rossby wave propagation
     & $\{(s,t)~|~ s\geq s_{{\rm tr}}\}$ \\ 
$D^{{\rm all}}$ &  entire domain & $D^{{\rm cv}}\cup D^{{\rm rw}}$ \\ 
$D^{[k]}$ & domain for the  $k$-th oscillation period
    & $\{(s,t)~|~ t\in T^{[k]}\}$ \\  
$D(t)$ & transient (with respect to present time $t$)  
    & $\{(s,\tau)~|~ \tau<t\}$ \\  
$D^{[k]}(t)$ & transient (with respect to present time $t$)
  during the $k$-th oscillation period in ${\bf u}'({\bf x},t)$
   & $\{(s,\tau)~|~ \tau<t \mbox{ and } \tau\in T^{[k]}\}$ \\  \hline
\end{tabular}
\caption{Definition of the sub-domains for the double-gyre
 application.}
\label{tbl:D}
\end{center}
\end{table}

\newpage
\addcontentsline{toc}{section}{\protect\numberline{}{Figures}}
\listoffigures
\clearpage
% figure: reference state of double-gyre - - - - -

% - - - - - - - - - - - - - - - - - - - - - - - - - - - - - -
% figure: must along entire C - - - - -

% figure: eddy of double-gyre - - - - -

% figure: mst,mpst and mfst for once cycle  - - - - -

% figure: mst,mpst and mfst for once cycle  - - - - -

% figure: m for all and cylce  - - - - -

\newpage

% figure: m for all and cylce  - - - - -

\end{document}